\documentclass[secnumarabic,amssymb, nobibnotes, prl, superscriptaddress, preprint]{revtex4-1}
\usepackage{comment}

\usepackage{graphicx}
\usepackage{dcolumn}
\usepackage{bm}
\usepackage{amssymb}

\usepackage{geometry}
\usepackage{longtable}
\usepackage{url}

\begin{document}

\title{\bf Electron-impact direct double ionization}%

\author{V.~Jonauskas}%
\email[]{Valdas.Jonauskas@tfai.vu.lt}
\affiliation{Institute of Theoretical Physics and Astronomy, Vilnius
University, \ A. Go\v{s}tauto 12, LT-01108  Vilnius, Lithuania}

\author{A.~Prancikevi\v{c}ius}
\affiliation{Leiden Institute of Physics, Leiden University, P.O. Box 9504, 2300 RA Leiden, The Netherlands}

\author{\v{S}.~Masys}
\affiliation{Institute of Theoretical Physics and Astronomy, Vilnius
University, \ A. Go\v{s}tauto 12, LT-01108  Vilnius, Lithuania}

\author{A.~Kynien\.{e}}
\affiliation{Institute of Theoretical Physics and Astronomy, Vilnius
University, \ A. Go\v{s}tauto 12, LT-01108  Vilnius, Lithuania}

\date{\today}%

\begin{abstract}

Electron-impact direct double ionization (DDI) process is studied as a sequence of  two and three step processes. Contribution from ionization-ionization, ionization-excitation-ionization, and excitation-ionization-ionization processes is taken into account. The present results help to resolve the long-standing discrepancies; in particular, a  good agreement with experimental measurements is obtained for double ionization cross-sections of $O^{1+}$, $O^{2+}$, $O^{3+}$, $C^{1+}$, and $Ar^{2+}$ ions. We show that distribution of the energy of scattered and ejected electrons, which participate in the next step of ionization, strongly affects DDI cross-sections.
\end{abstract}


\maketitle

Interactions of electrons with ions and atoms are atomic processes of fundamental nature. Electron-impact ionization plays a significant role not only in many fields of physics, but also in other sciences. Single ionization is usually the strongest among various ionization processes, but multiple ionization is important in various environments with an abundance of energetic electrons.
Compared with other multiple ionization processes, double ionization (DI) has the largest impact to ionization state distribution. 
Direct and indirect processes are responsible for the formation of the charge state of the resulting ion with two removed electrons. Indirect process is determined by ionization-autoionization (IA), while direct process  occurs due to the simultaneous ionization of two electrons in the target ion. 
From the theoretical point of view, the many-body problem with a few outgoing electrons in the vacuum has to be solved in the latter case. 

DI has been widely studied for the light elements theoretically and experimentally \cite{1991sscp_54_13_muller, 2003pr_374_91_berakdar, 2008aamop_55_293_muller, 2012epjd_66_1_colgan}. Many developed theoretical methods \cite{1973jpb_6_270_tweed, 1999jpb_32_5047_kheifets, 2000jpb_33_4323_defrance, 2004pra_70_032705_pindzola}   show a good agreement with experimental measurements for the two-electron systems. 
Time-dependent close-coupling approach has been used to analyze more complex systems with more than two electrons \cite{2009jpb_42_215204_pindzola, 2010jpb_43_105204_pindzola, 2011jpb_44_105202_pindzola}. However, those calculations are cumbersome even for the light atoms. 
From the perspective of classical approach, the mechanisms of DI were identified by Gryzinski \cite{1965pr_138_A336_gryzinski}. Unfortunately, this approach for one or another reason failed to provide a good agreements with measurements  in most cases.


The main aim of our paper is to show that direct double ionization (DDI) can be investigated as a sequence of a few processes which take place in the atomic system. 
To demonstrate possibilities of our approach, we have performed calculations of DI cross-sections for the light ions: $O^{1+}$, $O^{2+}$, $O^{3+}$, $C^{1+}$, and $Ar^{2+}$. In addition, DI cross-sections have been studied in $W^{5+}$ and $W^{25+}$ ions using Unresolved Transition Array (UTA) approach.

First of all, we consider an ensemble of ions or atoms which undergo collisions with electrons of energy $\varepsilon_0$. Some of the ions of the ensemble are excited to the higher levels while the others reach the next ionization stage after electron impact. After the first collision with electrons, populations of ions in various levels  are different because cross-sections of the electron-impact excitation and ionization processes to these levels are also different. We assume that after the first ionization process  from $i$ level  to $j$ level with cross-section $\sigma^{CI}_{ij}(\varepsilon_{0})$ the population of the final level is $p_{j}$.  An additional electron can be removed by scattered or ejected electrons. Probability to remove the additional electron from $nl$ shell, when atomic system undergoes a transition from $j$ level   to $f$ level, can be expressed by 
$
\sigma^{CI}_{jf} (\varepsilon_{1})/(4 \pi \bar{r}^{2}_{nl})
$ \cite{1965pr_138_A336_gryzinski}.
Here, $\varepsilon_{1}$ is the energy of the scattered or ejected electron, which removes the additional electron,  $\bar{r}_{nl}$ is the average distance among the electrons in the $nl$ shell. Assuming that density of electrons in the shell is uniform we can write $\bar{R}_{nl} \approx \bar{r}_{nl} N_{nl}^{1/3}$, where $\bar{R}_{nl}$ is the mean distance of the electrons from the nucleus, $N_{nl}$ is the number of electrons in the $nl$ shell. 

Thus, the equation for DDI from $i$ level to $j$ level through the ionization-ionization (II) path can be written as
\begin{equation}
\sigma^{DDI-II}_{if}(\varepsilon_{0}) = \sum_{j} \sigma^{CI}_{ij} (\varepsilon_{0}) p_{j} \frac{ \sigma^{CI}_{jf} (\varepsilon_{1}) (N_{nl})^{2/3}}{4 \pi \bar{R}^{2}_{nl}} .
\label{II}
\end{equation}

Additional intensive paths of DDI go through excitation-ionization-ionization (EII) and ionization-excitation-ionization (IEI) processes.  For EII process, the DDI cross-section can be expressed by the equation:
\[
\sigma^{DDI-EII}_{if}(\varepsilon_{0}) = \sum_{kj} \sigma^{CE}_{ik} (\varepsilon_{0}) p_{k} 
\]
\begin{equation}
\times \frac{ \sigma^{CI}_{kj}(\varepsilon_{1}) (N_{nl})^{2/3}}{4 \pi \bar{R}^{2}_{nl}}
\frac{ \sigma^{CI}_{jf}(\varepsilon_{2}) (N_{n'l'})^{2/3}}{4 \pi \bar{R}^{2}_{n'l'}} .
\label{EII}
\end{equation} 
Here, $p_{k}$ stands for the population of the excited $k$ state of the initial ion, $\varepsilon_{1} = \varepsilon_{0} - \Delta E_{ik}$, $\Delta E_{ik}$ is a transition energy,  $\varepsilon_{2}$ is the energy of scattered or ejected electron.

In a similar way, the DDI cross-section for IEI process can be expressed by the equation:
\[
\sigma^{DDI-IEI}_{if}(\varepsilon_{0}) = \sum_{kj} \sigma^{CI}_{ik} (\varepsilon_{0}) p_{k} 
\]
\begin{equation}
\times \frac{ \sigma^{CE}_{kj}(\varepsilon_{1}) (N_{nl})^{2/3}}{4 \pi \bar{R}^{2}_{nl}}
\frac{ \sigma^{CI}_{jf}(\varepsilon_{2}) (N_{n'l'})^{2/3}}{4 \pi \bar{R}^{2}_{n'l'}} .
\label{IEI}
\end{equation} 

It should be noted that the main difference between our method and the approach proposed by Gryzinski \cite{1965pr_138_A336_gryzinski} is the  population of levels included into Eqs. (\ref{II}), (\ref{EII}), and (\ref{IEI}). Moreover, previously not described  additional processes, such as EII and IEI, are also determined.

Flexible Atomic Code (FAC) \cite{2008cjp_86_675_Gu} is employed to obtain autoionization transition probabilities as well as electron-impact excitation and single ionization cross-sections in the distorted wave approximation. The largest uncertainties in the calculation of DDI cross-sections using Eqs. (\ref{II}), (\ref{EII}), and (\ref{IEI}) come from calculation of ionization cross-sections within the distorted wave approximation since it is not clear which mean configuration, i.e. of ionizing or ionized ion, has to be applied \cite{2008cjp_86_675_Gu}. Furthermore, the previous studies demonstrate that incident and scattered electron continuum orbitals can be evaluated in the potential of ionizing \cite{1981pra_24_1278_younger} or ionized \cite{1991jpb_24_l405_botero} ion. Therefore, we have checked which approach gives a better agreement with experimental measurements for single and double ionization cross-sections and included only those calculations.

\begin{figure}
 \includegraphics[scale=0.3]{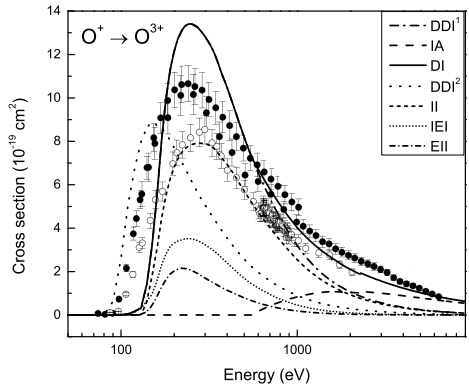}%
 \caption{\label{o1} Electron-impact DI cross-sections for $O^{1+}$. $\mathrm{DDI}^{1}$ stands for DDI cross-sections when scattered and ejected electrons share the excess energy,  $\mathrm{DDI}^{2}$ - one of the electrons takes all the available energy after ionization,  $\mathrm{DI}$ - sum of  $\mathrm{DDI}^{1}$ and  $\mathrm{IA}$ parts. See explanations in the text for the other processes. Experiment: solid circles \cite{1999ps_1999_285_westermann}, open circles \cite{1994jpb_27_2383_zambra}.}
 \end{figure}

The theoretical electron-impact DI cross-sections along with experimental values for $O^{1+}$ ion are displayed in Fig. \ref{o1}. The calculated DDI cross-sections correspond to the two cases of energy distribution of scattered and ejected electrons. In one case, the excess energy is taken by  one of the electrons participating in the collision. Only this electron participates in the further process which results in DI. In the other case, the ejected and scattered electrons  share the excess  energy. Further collisions of one of two available electrons with any target electrons can lead to the removal of the additional electron from the system. Theoretical cross-sections when electrons share the excess energy agree quite well with experimental values \cite{1999ps_1999_285_westermann} obtained in Giessen the electron-ion crossed-beam set-up \cite{1989jpb_22_1241_tinschert}. The later measurements from Zambra {\it et al.} \cite{1994jpb_27_2383_zambra} show about 20 \% smaller values than  \cite{1999ps_1999_285_westermann}. It can be explained by the fact that the different metastable fractions of $O^{1+}$ ion are present in the ion beams. Our data correspond to the ionization from the ground level. Calculations show that contributions from the higher levels of $2s^{2} 2p^{3}$ configuration can  have a decreasing or increasing character to the total cross-section. It is worth to note that at high energy limit electrons after collision tend to equally share  the excess energy. On the other hand, one of the electrons acquires large part of the excess energy at lower energies of the incident electron. It is obvious that the differences between  theoretical and experimental data can be removed by analyzing electron energy distribution after impact ionization. 

As it can be seen from Fig. \ref{o1}, the contribution from II process dominates over the contribution from IEI and EII processes. The cross-sections from IEI process are about 50 \% larger compared with the cross-sections from EII process. This can  be explained by the fact that initial ionization process is relatively stronger than excitation. Calculations show that the relative population of $2s^{2}2p^{2}$ configuration is equal to 30 \% and the population of $2s^{1}2p^{3}$ configuration amounts to 10 \%  at electron energy of 300 eV. Electron-impact excitation gives the relative population of 33 \% for $2s^{1}2p^{4}$ configuration at the same electron energy. 

\begin{figure}
 \includegraphics[scale=0.3]{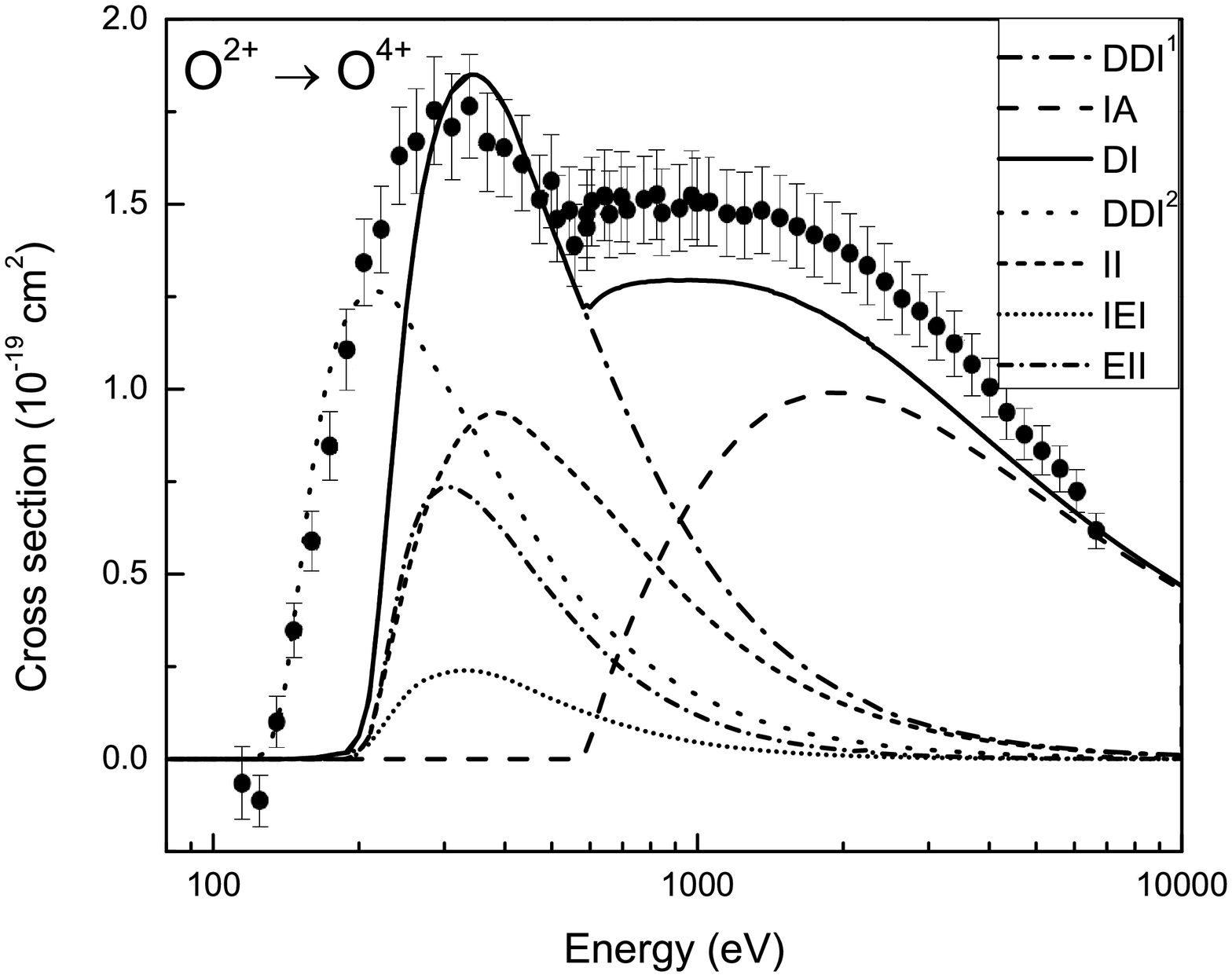}%
 \caption{\label{o2}  Same as Fig. \ref{o1} but for $O^{2+}$. }
 \end{figure}

 The cross-sections show a well-distinguished two-maxima structure in the case of $O^{2+}$ ion (Fig. \ref{o2}). Theoretical cross-sections correspond to the ionization from the excited $2p_{0.5}2p_{1.5} (J=2)$ level of the ion. DDI cross-sections from the other levels of the ground configuration give smaller or larger values. On the other hand, the indirect part is not strongly affected by which level is used to calculate cross-sections. At low incident electron energies, a better agreement with the experimental cross-sections is obtained if after impact, which leads to the ionization, one of the electrons takes all the excess energy. And again, the largest contribution to DDI comes from II process. However, the contribution of IEI process is relatively larger compared to the $O^{1+}$ case. The analysis of population of configurations after the first collision, which leads to excitation or ionization, shows that the largest flux goes to the excited configurations of the initial ion. 
 
 There are some differences between theoretical and experimental data at electron energies where direct and indirect processes start to overlap. This disagreement can be explained by the additional contribution from the excited $2s2p^{3}$ configuration. Our calculations show that cross-section maximum for the direct part reaches $3\cdot 10^{-19}$ cm$^{2}$ at 360 eV electron energy for this configuration. 
 

\begin{figure}
 \includegraphics[scale=0.3]{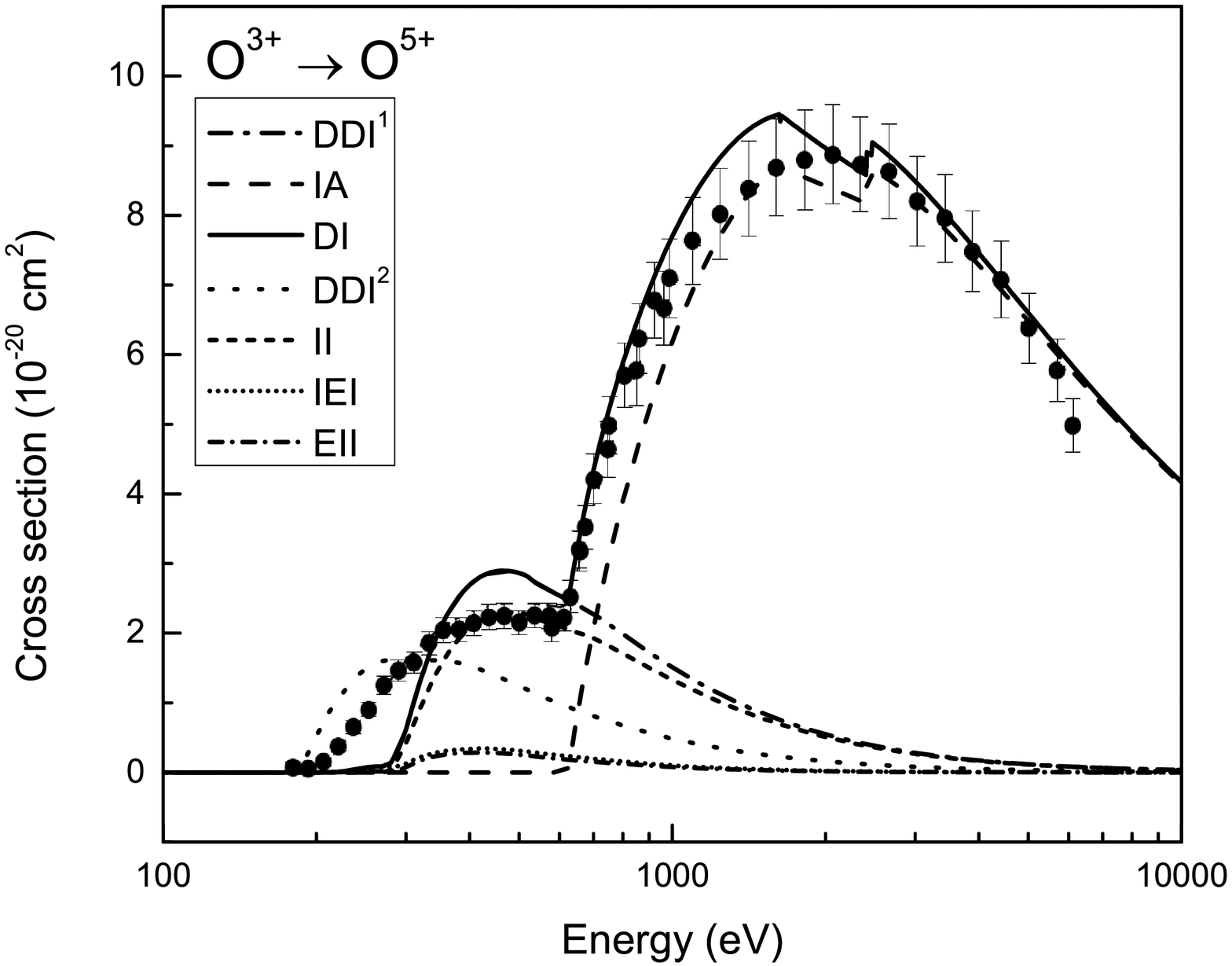}%
 \caption{\label{o3} Same as Fig. \ref{o1} but for $O^{3+}$. }
 \end{figure}

For $O^{3+}$ ion, the indirect part of DI cross-section dominates over the direct part (Fig. \ref{o3}). Calculations correspond to the ionization from the lowest level of the first excited $2s^{1}2p^{1}$ configuration. The contribution of the ground configuration to the direct part is about 40 \% smaller. At low impact energies, a better agreement of theoretical values with experimental ones is obtained when one of the electrons takes all the available energy.  The difference between the theoretical and experimental cross-sections can be related to the different electron energy distribution after collision. 
 
\begin{figure}
 \includegraphics[scale=0.3]{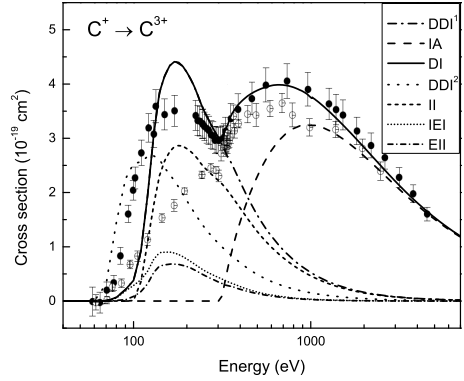}%
 \caption{\label{c1}  Same as Fig. \ref{o1} but for $C^{1+}$. }
 \end{figure}

In the case of $C^{1+}$ ion, a two-maxima structure is also seen for DI cross-sections (Fig. \ref{c1}). We present  cross-sections  for the lowest level of the first excited  $2s^{1}2p^{2}$ configuration. The cross-sections from the ground level have slightly lower values compared with ionization from the level of the excited configuration. On the other hand, the cross-sections from the other two long-lived levels of the  $2s^{1}2p^{2}$ configuration are higher or lower than from the lowest level of the configuration. 

\begin{figure}
 \includegraphics[scale=0.3]{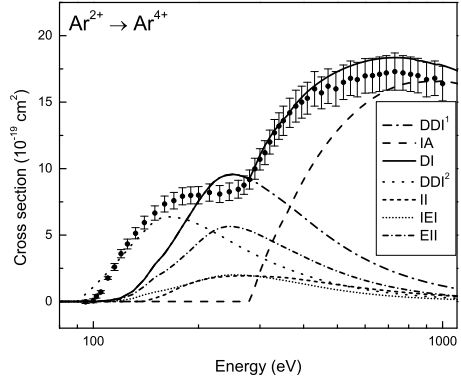}%
 \caption{\label{ar2}  Same as Fig. \ref{o1} but for $Ar^{2+}$. Solid circles - experiment \cite{1989jpb_22_1241_tinschert}.}
 \end{figure}

The ground configuration of $Ar^{2+}$ ion has filled shells up to $3s$ with valence electrons in $3p$ shell. Calculated electron-impact DI cross-sections from the excited $3p_{1.5}^2 (J=0)$ level of the ground configuration are displayed in Fig. \ref{ar2} along with the experimental values \cite{1989jpb_22_1241_tinschert}. A better agreement with the experiment at lower electron energies is obtained if one assumes that one of electrons after ionization takes all the available energy. A share of the excess energy starts to dominate near the peak of DDI cross-section. However, many long-lived levels of the first excited $3p^{3}3d$ configuration have to be studied to find out the metastable fraction in the ion source. 

Encouraged by the obtained results, we also investigated  DI cross sections for $W^{5+}$ and $W^{25+}$ ions using UTA approach. Comparison with experimental data for   $W^{5+}$ ion shows the same tendencies for distribution of electron energies as for the other studied  ions - one of the electrons takes main part of the available energy of the collision at low energies. However, large number of long-lived levels of $4f^{13}5d^{2}$, $5p^{5}5d^{2}$, and $6s$ configurations have to be analyzed in order to estimate their contribution to DI cross-sections. DDI cross-sections for $W^{25+}$ ion are two orders lower than contribution from IA process. It confirms the fact that for highly charged ions DDI influence is very small.

To conclude, here we have developed a method that considers electron-impact DDI process as a consequence of two and three step processes. Excitation and ionization processes after collision of incident electron with target are studied to obtain population of levels for further steps. We have demonstrated that the  method can be easily applied for complex ions. Much work still needs to be done analyzing distribution of electron energies after the first ionization process.

This research was funded by European Social Fund under the Global Grant Measure (No.: VP1-3.1-\v{S}MM-07-K-02-015). Part of computations were performed on resources at the High Performance Computing Center „HPC Sauletekis“ in Vilnius University Faculty of Physics.


%

\end{document}